\documentclass[aps,pra,showpacs,twocolumn,superscriptaddress]{revtex4}
\usepackage{amsmath,graphicx,epsfig,color}
\usepackage{amssymb,graphics,psfrag}
\usepackage[hypertex]{hyperref}

\newcommand{\ket}[1]{|#1\rangle}
\newcommand{\bra}[1]{\langle#1|}
\newcommand{\ketbra}[1]{\ket{#1}\bra{#1}}
\newcommand{\tr}{\text{tr}}
\newcommand{\ten}{\otimes}

\newcommand{\sA}{s_a}
\newcommand{\sB}{s_b}
\newcommand{\oA}{o_a}
\newcommand{\oB}{o_b}

\newcommand{\idx}[1]{#1\text{-th}}
\newcommand{\ii}{~{\rm i}}
\renewcommand{\exp}{{\rm e}}
\newcommand{\I}{\mathcal{I}}
\def\A{\mathcal{A}}
\def\B{\mathcal{B}}
\def\M{\mathcal{M}}
\def\S{\mathcal{S}}
\def\Sj{\S_{\mbox{\tiny Ji}}}

\newcommand{\ProbTwGJ}{p^{\oA\oB}_{\A\B}(\sA,\sB)}
\newcommand{\ProbTwGMA}{p^{\oA}_{\A}(\sA)}
\newcommand{\ProbTwGMB}{p^{\oB}_{\B}(\sB)}

\newcommand{\POVMA}[2]{A_{#1}^{#2}}
\newcommand{\POVMB}[2]{B_{#1}^{#2}}
\newcommand{\Bell}{\mathcal{B}}
\newcommand{\SqmBI}[1]{\mathcal{S}_{\mbox{\tiny QM}}^{\mbox{\tiny (#1)}}}
\newcommand{\SqmBIMP}[1]{\mathcal{S}_{\mbox{\tiny QM, mp}}^{\mbox{\tiny (#1)}}}

\newcommand{\DeltaMax}{\Delta_{\mbox{\tiny max}}^{(d)}}
\newcommand{\DeltaLB}{\Delta_{\mbox{\tiny LB}}^{(d)}}

\newcommand{\DeltaMin}{\Delta_{\mbox{\tiny min}}^{(d)}}
\newcommand{\Idp}{I^+_d}
\newcommand{\ibid}{{\em ibid.~}}
\newcommand{\qic}{Quant. Inf. Comput. }
\newcommand{\fp}{Found. Phys. }
\newcommand{\njp}{New J. Phys. }

\begin{document}

\title{Reexamination of a multisetting Bell inequality for qudits}

\author{Yeong-Cherng~Liang}
\email{ycliang@physics.usyd.edu.au} \affiliation{School of
Physics, The University of Sydney, New South Wales 2006,
Australia.}

\author{Chu-Wee Lim}
\email{lchuwee@dso.org.sg} \affiliation{DSO National
Laboratories, 20 Science Park Drive, Singapore 118230.}

\author{Dong-Ling Deng}
\affiliation{Theoretical Physics Division, Chern Institute of
Mathematics, Nankai University, Tianjin 300071, People's~Republic of
China.}

\date{\today}

\begin{abstract}
The class of $d$-setting, $d$-outcome Bell inequalities proposed by
Ji and collaborators [\pra {\bf 78}, 052103] are reexamined. For
every positive integer $d>2$, we show that the corresponding
non-trivial Bell inequality for probabilities provides the maximum
classical winning probability of the Clauser-Horne-Shimony-Holt-like
game with $d$ inputs and $d$ outputs. We also demonstrate that the
general classical upper bounds given by Ji {\em et al.} are
underestimated, which invalidates many of the corresponding
correlation inequalities presented thereof. We remedy this problem,
partially, by providing the actual classical upper bound for $d\le
13$ (including non-prime values of $d$). We further determine that
for prime value $d$ in this range, most of these probability and
correlation inequalities are tight, i.e., facet-inducing for the
respective classical correlation polytope. Stronger lower and upper
bounds on the quantum violation of these inequalities are obtained.
In particular, we prove that once the probability inequalities are
given, their correlation counterparts given by Ji and co-workers are
no longer relevant in terms of detecting the entanglement of a
quantum state.
\end{abstract}

\maketitle

\section{Introduction}

Bell inequalities~\cite{J.S.Bell:1964,R.F.Werner:QIC:2001}, being
constraints that have to be satisfied by classical correlations, have
long played an important role in shaping our current world
view~\cite{N.D.Mermin:PhysToday:1994}. With the advent of quantum
information science, these inequalities have also found applications
in the design of quantum key distribution
protocol~\cite{A.K.Ekert:PRL:1991} and its security
analysis~\cite{QKD:Security}, as well as the reduction of
communication complexity~\cite{C.Brukner:PRL:2004}. More recently,
there is also a growing interest in thinking about these inequalities
in the form of non-local games~\cite{R.Cleve:0404076:2004,
A.C.Doherty:QMP, J.Silman:PLA:3796} which, in turn, are closely
related to the studies of interactive proof systems in computer
science (see, for example, Refs.~\cite{R.Cleve:0404076:2004}
and~\cite{A.C.Doherty:QMP}).

To date, the studies of Bell inequalities have focused predominantly
on those involving only binary outcomes, such as the
Bell-Clauser-Horne-Shimony-Holt
inequality~\cite{CHSH,J.S.Bell:Speakable} and the Bell-Clauser Horne
(henceforth abbreviated, respectively, as Bell-CHSH and Bell-CH)
inequality~\cite{CH:PRD:1974,fn:two-outcome} (see, for example,
Refs.~\cite{D.Collins:JPA:2004, T.Ito:PRA:2006, N.Brunner:PLA:2008,
K.F.Pal:PRA:022120, Y.C.Liang:Thesis} and references therein for a
review on bipartite two-outcome inequalities). This is, of course, by
no means accidental as many of the quantum information processing
protocols have been developed explicitly with qubits, i.e., two-level
quantum systems in mind~\cite{Book:MikeIke}. However, given that
higher-dimensional quantum systems are gaining importance in quantum
information processing
tasks~\cite{T.C.Ralph:PRA:022313,B.P.Lanyon:NatPhys:134}, the time is
now ripe to also perform further studies on multiple-outcome Bell
inequalities, which are naturally suited for higher-dimensional
quantum systems.

In this regard, we note that there are only relatively few works
devoted to the studies of such Bell inequalities and their
quantum-mechanical violations. For an experimental scenario involving
only two subsystems, the pioneering work by Collins~{\em et
al.}~\cite{D.Collins:PRL:2002} resulted in a class of Bell
inequalities that involves two multiple-outcome measurements per site
(see also Refs.~\cite{D.Kaszlikowski:PRA:2002}
and~\cite{S.Zohren:PRL:2008}). This class of inequalities, now known
as the Collins-Gisin-Linden-Massar-Popescu (CGLMP) inequalities, is
tight~\cite{Ll.Masanes:QIC:2003}, i.e., they represent boundaries of
the corresponding set of classical correlations, or more precisely,
facets of the respective correlation polytope~\cite{CorPolytope} (for
a review on the subject of polytope, see Ref.~\cite{Polytopes}).

Apart from the CGLMP inequalities, there are only a few other classes
of Bell inequalities that are  specifically catered for multiple
outcomes. Some of these are defined in terms of joint and marginal
probabilities of experimental
outcomes~\cite{H.Bechmann-Pasquinucci:PRA:2003, D.Collins:JPA:2004,
A.Acin:PRL:250404, H.Buhrman:PRA:2005, J.L.Chen:PRA78:032107,
S-W.Lee:0803.3097, D.L.Deng:0902.3534}, whereas the
others~\cite{J.L.Chen:PRA71:032107, W.Son,
L.B.Fu:PRL:2004,Ji:PRA:052103,J.L.Chen:PRA:012111} are defined in terms
of correlation functions --- i.e., expectation value of the product of
experimental outcomes. In general, however, very little is known about
the tightness of these inequalities~\cite{D.Collins:JPA:2004,
S-W.Lee:PRA:2007, J.L.Chen:PRA78:032107, S-W.Lee:0803.3097,
D.L.Deng:0902.3534}.

An interesting feature of tight multiple-outcome Bell inequalities is
that, except for those introduced in Ref.~\cite{S-W.Lee:0803.3097},
they are typically violated maximally by non-maximally entangled
states~\cite{A.Acin:PRA:052325, J.L.Chen:PRA:032106, N.Gisin:0702021,
D.L.Deng:0902.3534}. This and other evidences gathered from the studies
of non-local apparatuses~\cite{S.Popescu:FP:379} --- collectively known
as ``an anomaly of non-locality"~\cite{A.A.Methot:QIC:2007} --- have
led to the proposal of seeing quantum entanglement and Bell inequality
violation as fundamentally different
resources~\cite{N.Brunner:NJP:2005}, even though we now know that all
bipartite entangled states cannot be simulated by classical
correlations alone~\cite{Ll.Masanes:PRL:2008}.

In this paper, we reexamine the class of bipartite $d$-setting,
$d$-outcome Bell correlation inequalities proposed by Ji {\em
et~al.}~\cite{Ji:PRA:052103}. In Sec.~\ref{Sec:ClassicalBound}, we
rewrite these correlation inequalities as Bell inequalities for
probabilities and show that it admits a natural interpretation within
the framework of the so-called CHSH game~\cite{R.Cleve:0404076:2004}
(but now with $d$ inputs and $d$ outputs). In the same section, we
provide, for $d\le13$, the actual classical upper bound and for the
more complicated scenarios, some non-trivial estimates thereof. The
tightness of these inequalities is discussed in
Section~\ref{Sec:Tightness}. After that, in
Sec.~\ref{Sec:QuantumViolation}, we investigate the quantum violation
of the probabilities inequalities and compare them against those
obtained in Ref.~\cite{Ji:PRA:052103} using their correlation
counterparts. We will conclude with a summary of results and some
possibilities for future research in Sec.~\ref{Sec:Conclusion}.
Throughout, our discussion focuses on the scenarios where $d$ is a
prime number; the analogous computational results for non-prime value
of $d$ (with $d\le12$) are summarized briefly in
Appendix~\ref{App:non-prime}.

\section{The Bell functions and their classical bounds}
\label{Sec:ClassicalBound}

The Bell inequalities proposed by Ji and
co-workers~\cite{Ji:PRA:052103} are applicable to an experimental
scenario where two spatially separated experimenters (hereafter
called Alice and Bob) are each allowed to perform $d$ alternative
measurements, with $d$ being arbitrary {\em prime} number.
Specifically, if we denote by $\omega=\exp^{\ii2\pi/d}$ the $\idx{d}$
root of unity, Ji {\em et~al.} consider local observables $A_{\sA}$
and $B_{\sB}$ that are unitary so that each measurement admits the
$d$ possible outcomes $\{\omega^k\}_{k=0}^{d-1}$. In these notations,
the Bell function and the correlation inequalities presented in
Ref.~\cite{Ji:PRA:052103}
--- up to a factor of $1/(d-1)$ --- read as
\begin{subequations}\label{Eq:Ji}
\begin{gather}\label{Eq:BellFn:Ji}
    \Sj=\sum_{n=1}^{d-1}\sum_{\sA,\sB=0}^{d-1} \omega^{n\sA\sB}\langle
    \left(A_{\sA}\right)^n \left(B_{\sB}\right)^n\rangle,\\
    \label{Ineq:Ji}
    d\,\DeltaMin \le \Sj +d^2 \le d\,\DeltaMax,
\end{gather}
where the classical upper (lower) bound is determined by maximizing
(minimizing) over all extremal (deterministic) classical strategies
$\oA(\sA)$ and $\oB(\sB)$, i.e.,
\begin{align}
    \label{Eq:DeltaMax}
    &\DeltaMax\equiv \max \sum_{\sA,\sB}
    \delta_{\sA\sB+\oA(\sA) +\oB(\sB)},\\
    \label{Eq:DeltaMin}\text{and}\quad
    &\DeltaMin\equiv \min \sum_{\sA,\sB}
    \delta_{\sA\sB+\oA(\sA) +\oB(\sB)},
\end{align}
\end{subequations}
and here, $\delta_{j}$ is a shorthand for the Kronecker delta
$\delta_{0,\, j\!\!\mod\! d}$. In Eq.~\eqref{Eq:BellFn:Ji}, $\langle
\left(A_{\sA}\right)^n \left(B_{\sB}\right)^n\rangle$ is a {\em
correlation function} that gives the statistical average of the
product of measurement outcomes of $\left(A_{\sA}\right)^n$ and
$\left(B_{\sB}\right)^n$. Hereafter, we shall refer to the inequality
upper (lower) bounding $\Sj$ in Eq.~\eqref{Ineq:Ji} as $I_{c,d}^{+}$
($I_{c,d}^{-}$).

For any given setup of the Bell experiment, the set of correlation
functions $\left\{\langle \left(A_{\sA}\right)^n
\left(B_{\sB}\right)^n\rangle\right\}_{\sA,\sB,n}$ can be collected
together and written as the entries of a vector in
$\mathbb{C}^{d^2(d-1)}$. It is known that the set of such vectors
allowed by a local hidden-variable theory (LHVT) forms a convex
polytope~\cite{CorPolytope}
--- i.e., {\em loosely}, higher-dimensional generalizations of convex
polygons --- called a classical {\em correlation polytope}. Each
$I_{c,d}^{\pm}$ given in Eq.~\eqref{Eq:Ji} then defines a hyperplane in
the space of complex correlations $\mathbb{C}^{d^2(d-1)}$, separating
(some) correlations not attainable using LHVT from the classical
correlation polytope.

Now, let us further denote by $\ProbTwGJ$ the joint probability of
Alice observing the $\idx{\oA}$ outcome and Bob observing the
$\idx{\oB}$ outcome conditioned on her measuring $A_{\sA}$ and him
measuring $B_{\sB}$; likewise for the marginal probabilities
$\ProbTwGMA$ and $\ProbTwGMB$. From Eq.~\eqref{Eq:Ji} and the fact that
{\em classically},
\begin{equation}\label{Eq:Mapping}
    \langle\left(A_{\sA}\right)^n \left(B_{\sB}\right)^n\rangle
    = \sum_{\oA=0}^{d-1}\sum_{\oB=0}^{d-1} \omega^{n\oA+n\oB} \ProbTwGJ,
\end{equation}
it can be shown that the following Bell function,
\begin{subequations}\label{Ineq:Ipm}
\begin{equation}\label{Eq:BellFn:Prob}
    \S =\frac{1}{d^2}
    \sum_{\sA,\sB,\oA,\oB=0}^{d-1}\delta_{\sA\sB+\oA +\oB}\,\ProbTwGJ,
\end{equation}
must also be bounded from below and above as follows:
\begin{equation}
    I^-_d: \frac{1}{d^2}\DeltaMin\le\S,\quad
    I^+_d: \S\le \frac{1}{d^2}\DeltaMax.
\end{equation}
\end{subequations}
This gives rise to two classes of linear Bell inequalities for
probabilities.

A few remarks are now in order. In contrast with $I_{c,d}^{\pm}$ given
in Eq.~\eqref{Eq:Ji}, the Bell inequalities given in
Eq.~\eqref{Ineq:Ipm} live in the space of real correlations
$\mathbb{R}^{d^4}$ where each vector in the space has entries given by
all the $d^4$ distinct joint probabilities $\ProbTwGJ$. Moreover, it is
also easy to see that the requirement that each experimenter must
perform unitary measurements in Eq.~\eqref{Eq:Ji} is now lifted from
$I_d^{\pm}$; Alice and Bob are free to assign arbitrary values to their
measurement outcomes.

On the other hand, note that $\S$ only involves non-negative
combination of $\ProbTwGJ$ and that the right-hand-side of
Eq.~\eqref{Eq:BellFn:Prob} is upper bounded~\cite{fn:ub} by 1. As a
result, $\S$ can also be seen as the winning probability of a
two-prover, one-round {\em unique game}~\cite{S.Khot:ACM:2002} whereby
Alice and Bob win if and only if the answers that they provide
$\oA(\sA)$ and $\oB(\sB)$ for the questions $\sA$, $\sB$ (supplied to
them with uniform probability) are such that:
\begin{equation}\label{Eq:WinningStrategy}
    \sA\sB+\oA(\sA)+\oB(\sB)\mod d=0.
\end{equation}
This is clearly a direct generalization of the CHSH game presented in
Ref.~\cite{R.Cleve:0404076:2004}. Classically, the winning probability
of the CHSH game [corresponding to $d=2$ in
Eq.~\eqref{Eq:WinningStrategy}] is upper bounded by $3/4$, but one can
easily check that this is just the requirement of the Bell-CH
inequalities~\cite{CH:PRD:1974}. For the rest of the paper, we will
thus focus on scenarios where $d>2$ and analyze the probability
inequalities $I_d^{\pm}$ in connection with their correlation
counterpart $I_{c,d}^{\pm}$.

What are the actual classical bounds for these Bell inequalities? Here,
we follow Ref.~\cite{Ji:PRA:052103} and consider a $d\times d$ matrix
$\mathcal{M}_{d}$ with its $(\sA+1, \sB+1)$ matrix element given by the
left-hand-side of Eq.~\eqref{Eq:WinningStrategy}. A given extremal
classical strategy, i.e., one that satisfies,
\begin{subequations}\label{Eq:ExtremalProb}
\begin{gather}\label{Eq:Factorization}
    \ProbTwGJ=\ProbTwGMA\ProbTwGMB,\\
    \label{Eq:Prob=01}
    \ProbTwGMA=0,1,\quad \ProbTwGMB=0,1,
\end{gather}
\end{subequations}
then gives rise to a classical value of $\S$ and $\Sj$ determined by
the number of zero entries in the corresponding matrix $\M_d$. For
$d>2$, the following classical strategy~\cite{Ji:PRA:052103}
\begin{gather*}
    \oA(\sA)=\sA-1~\forall~\sA\neq 0,\quad \oA(0)=0;\\
    \oB(\sB)=1~\forall~\sB\le d-2,\quad \oB(d-1)=2
\end{gather*}
gives $\S=0$. This, together with the non-negativity of $\S$ [cf.
Eq.~\eqref{Eq:BellFn:Prob}] show that $\DeltaMin=0$. Thus, $I_d^-$ and
$I_{c,d}^-$ for $d>2$ are Bell inequalities that are trivially
satisfied by any theories that respect the non-negativity of
probabilities.

As for the classical upper bound, it was estimated in
Ref.~\cite{Ji:PRA:052103} to be $\DeltaMax=3(d-1)$. While their
explicit analysis for $d=3$ is valid, it can be verified that the
following classical strategy for prime value $d>5$,
\begin{gather}
    \oA(\sA)=d-1~\forall~\sA\neq 0,1,\frac{d+1}{2},\nonumber\\
    \oA(0)=0,\quad \oA(1)=d-4,\quad \oA\left(\frac{d+1}{2}\right)
    =d-3,\nonumber\\
    \oB(\sB)=0~\forall~\sB> 2,\nonumber\\
    \oB(0)=1,\quad \oB(1)=3,\quad \oB(2)=2
    \label{Eq:Strategy:Cl:3d-2}
\end{gather}
gives $3d-2$ zero entries in $\M_d$, indicating that
$\DeltaMax\ge3d-2$.

In this regard, we note that the actual value of $\DeltaMax$ for
$d\le13$ can be determined by exhaustively searching through all
(extremal) classical strategies with the help from the following
observations (all arithmetic operations described below are to be
evaluated modulo $d$):
\begin{enumerate}
\item $\S$ is invariant under the simultaneous
    transformations: $\oA(\sA)\to\oA(\sA)+k$,
    $\oB(\sB)\to\oB(\sB)-k$ for all $\sA$, $\sB$ and all
    $k\in\{0,1,\ldots,d-1\}$. Without loss of generality,
    we may thus set $\oA(0)=0$.
\item For all $k\in\{1,2,\ldots,d-1\}$, the strategies
    $\{\oA(\sA),\oB(\sB)\}_{\sA,\sB=0}^{d-1}$ and
    $\{\oA'(\sA),\oB'(k\sB)\}_{\sA,\sB=0}^{d-1}$ give the same $\S$
    if $\oA'(\sA)=k\oA(\sA)$ and $\oB'(k\sB)=k\oB(\sB)$. This
    follows from $\sA(k\sB)+\oA'(\sA)+\oB'(k\sB)
    =k[\sA\sB+\oA(\sA)+\oB(\sB)]$ and thus the two strategies give
    the same number of zeros in the corresponding $d\times d$
    matrix $\mathcal{M}_d$. As a result, it suffices to consider
    $\oA(1)=0$ and $\oA(1)=1$ once we have set $\oA(0)=0$.
\item For a given $\sB$ and a given choice of
    $\{\oA(\sA)\}_{\sA=0}^{d-1}$, let $k$ be the number in
    $\{0,1,\ldots,d-1\}$ that occurs most frequently in the
    expression ``$\sA\sB+\oA(\sA)\,\mod{d}$" as $\sA$ varies from
    $0$ to $d-1$. The optimum choice of $\oB(\sB)$ for the given
    $\sB$ is $\oB(\sB)=d-k$.
\end{enumerate}
Explicit value of these $\DeltaMax$ can be found in
Table~\ref{tbl:DeltaMax}.

\begin{table}[!h]
\caption{\label{tbl:DeltaMax} $\DeltaMax$ and its lower bounds. The
first row gives the values of the parameter $d$. The second row gives
the best lower bound on $\DeltaMax$ that we have found whereas its
actual value, if known, is included in the third row of the table.}
\begin{ruledtabular}
\begin{tabular}{c|cccccccccc}
$d$         & 3 &  5 &  7 & 11 & 13 & 17 & 19 & 23 & 29 & 31\\\hline
$\DeltaLB$  & 6 & 12 & 19 & 37 & 47 & 66 & 79 & 99 & 135 & 148\\
$\DeltaMax$ & 6 & 12 & 19 & 37 & 47 & - & - & - & - & -\\
\end{tabular}
\end{ruledtabular}
\end{table}

For $d>13$, it seems formidable to search through all inequivalent
(extremal) classical strategies~\cite{fn:AllClassicalStrategy};
neither is the classical strategy given in
Eq.~\eqref{Eq:Strategy:Cl:3d-2} optimal. However, non-trivial lower
bounds on $\DeltaMax$ can be obtained by optimizing the classical
strategies of Alice $\{\oA(\sA)\}_{\sA=0}^{d-1}$ and Bob
$\{\oB(\sB)\}_{\sB=0}^{d-1}$ iteratively. Specifically, if we start
with a random choice of classical strategy for Alice, the optimal
strategy for Bob can be decided using the third observation mentioned
above. With this optimized classical strategy for Bob, we can in turn
determine the optimal classical strategy for Alice and so on and so
forth. The explicit values for some of these lower bounds, which we
will denote by $\DeltaLB$ can be found in Table~\ref{tbl:DeltaMax}.

\section{Tightness of Bell inequalities}\label{Sec:Tightness}

\begin{figure}[h!]
\scalebox{0.65}{\includegraphics{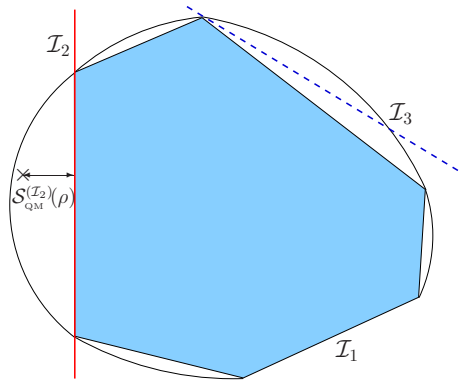}}
\caption{\label{Fig:Polytope}
(Color online) Schematic diagram of a two-dimensional plane in the space
of (quantum) correlations. The shaded (light blue) polygon only
consists of classical correlations whereas the convex region marked by
a circumscribing solid curve also consists of nonclassical
correlations. $\I_1$ corresponds to a trivial Bell inequality that
cannot be violated by quantum mechanics. The analog of a {\em tight}
Bell inequality, such as $I_2^+$ is the hyperplane given by $\I_2$ (red
solid line) whereas the analog of a non-tight Bell inequality such as
$I_3^+$ is given by $\I_3$ (blue dashed line). A quantum correlation
``$\times$" violates a Bell inequality if and only if the corresponding
hyperplane (eg. $\I_2$) separates ``$\times$" from the set of classical
correlations. }
\end{figure}

A natural question that follows is whether the Bell inequalities
$I_{c,d}^+$ and $\Idp$ are tight, or so called
facet-inducing~\cite{D.Avis:JPA:2005} for the respective set of
classical correlations. By analyzing the tightness of these
inequalities, we can gain insights on the structure of the
corresponding set of classical correlations
(Fig.~\ref{Fig:Polytope}). To this end, we note that the relevant
classical correlation polytope for $\Idp$ resides in a subspace of
$\mathbb{R}^{d^4}$ of dimension~\cite{D.Collins:JPA:2004}
$d_p=d^2(d-1)^2+2d(d-1)$. A Bell inequality is facet-inducing if and
only if the number of linearly independent extremal classical
probability (correlation) vectors saturating the inequality equals to
the dimension of the polytope. For $\Idp$, this can be shown to be
$d_p$ following a similar argument as that presented in
Ref.~\cite{Ll.Masanes:QIC:2003}; likewise for $I_{c,d}^+$, which can
be shown to be $d_c=d^2(d-1)$.

In this regard, we note that our investigation shows that
(Table~\ref{tbl:No.Saturating}) for $d\le 13$, most of these
probability and correlation inequalities are indeed facet-inducing.
Note, however, that there is {\em a priori} no reason to expect that
the Bell function given by Eq.~\eqref{Eq:BellFn:Ji} or
Eq.~\eqref{Eq:BellFn:Prob} would give rise to any tight Bell
inequalities.

\begin{table}[h!]
\caption{\label{tbl:No.Saturating} Computational results for the
tightness of $I_{c,d}^+$ and $I_d^+$ for $d=3,5,7,11$ and 13. The
first column of the table gives the parameter $d$. From the second to
the fourth columns, we have, respectively, the dimension of the
correlation polytope relevant to $I_{c,d}^+$, the number of {\em
linearly independent} extremal classical correlation vectors
saturating inequality $I_{c,d}^+$, and the tightness of $I_{c,d}^+$.
The analogous results for $\Idp$, when available, are presented from
the fifth to the seventh columns.}
\begin{ruledtabular}
\begin{tabular}{r||ccc||ccc}
$d$ & $d_c$ & $r^c_+$ & $I_{c,d}^+$ & $d_p$ & $r^p_+$ & $\Idp$
\\ \hline
  3  &    18 &    6 & non-tight &    48 &    18 & non-tight \\
  5  &   100 &  100 & tight     &   440 &   440 & tight \\
  7  &   294 &  294 & tight     &  1848 &  1848 & tight \\
  11 &  1210 & 1210 & tight     & 12320 & 12320 & tight\\
  13 &  2028 & 2028 & tight     & 24648 &     - & - \\
\end{tabular}
\end{ruledtabular}
\end{table}

\section{Quantum Violations}
\label{Sec:QuantumViolation}

In this section, we will investigate the quantum violations of $I_d^+$
and compare them against those presented in Ref.~\cite{Ji:PRA:052103}.
These quantities put bounds on the set of quantum
correlations~\cite{B.S.Cirelson:LMP:1980}. In particular, the maximal
violation of a {\em tight} Bell inequality for a given state $\rho$ is
a primitive measure of the extent to which $\rho$ is nonclassical. For
example, in Fig.~\ref{Fig:Polytope}, if a Bell inequality is such that
its Bell function gives zero for all points lying on $\I_2$, then the
maximal extent to which $\rho$ violates this inequality, denoted by
$\S^{(\mathcal{I}_2)}_{\mbox{\tiny QM}}(\rho)$, indicates the largest
possible distance between an $\I_2$-violating correlation derivable
from $\rho$ and the hyperplane $\I_2$. Likewise, the largest possible
distance between any point on the arc opposite to the polygon and
$\I_2$ gives rise to the maximal possible quantum violation of the Bell
inequality corresponding to $\I_2$.

Now, let us start by comparing the strength of $\Idp$ against
$I_{c,d}^+$ in terms of detecting nonclassical correlations present in
an entangled state. To this end, it is worth noting that if we denote
by $\POVMA{\sA}{\oA}$ the positive-operator-valued measure (POVM)
element associated with the $\idx{\oA}$ outcome of Alice's $\idx{\sA}$
measurement, the expression
\begin{equation}\label{Eq:AsA:nthPower}
    (A_{\sA})^n=\left[\sum_{\oA=0}^{d-1} \omega^{\oA}\POVMA{\sA}{\oA}\right]^n
    =\sum_{\oA=0}^{d-1} \omega^{n\oA}\POVMA{\sA}{\oA}
\end{equation}
holds true if and only if all the POVM elements satisfy
$\POVMA{\sA}{\oA}\POVMA{\sA}{\oA'}=\delta_{\oA\,\oA'}\POVMA{\sA}{\oA}$.
This implies, in particular, that in quantum mechanics,
Eq.~\eqref{Eq:Mapping} is only applicable when we are considering
projective measurements. In this case, it is easy to show using
Born's rule, Eqs.~\eqref{Eq:BellFn:Ji} and~\eqref{Eq:BellFn:Prob},
that for any quantum state $\rho$, the quantum values of their Bell
functions are related by
\begin{equation}\label{Eq:QuantumBellValues}
 \tr(\rho\,\Bell_{\mbox{\tiny Ji}})= d^3\tr\left(\rho\,\Bell\right)-d^2,
\end{equation}
where $\Bell_{\mbox{\tiny Ji}}$ and $\Bell$ are, respectively, the
Bell operator~\cite{S.L.Braunstein:PRL:1992} constructed from the
Bell inequalities $I_{c,d}^+$ and  $\Idp$. For generalized
measurements where Eq.~\eqref{Eq:AsA:nthPower} does not hold,
measuring $A_{\sA}$ no longer measures $(A_{\sA})^n$ concurrently;
$\Bell_{\mbox{\tiny Ji}}$ is also generally non-Hermitian in this
case. Clearly, this makes a test of the quantum mechanical prediction
against its classical counterpart [cf. Eq.~\eqref{Ineq:Ji}]
meaningless~\cite{fn:POVM}. Thus, any quantum state that violates the
inequalities presented in Ref.~\cite{Ji:PRA:052103} (and hence
$I_{c,d}^+$) must also violate $\Idp$ but the converse is not
necessarily true.

Numerically, by maximizing over the set of rank-1 projective
measurements realizable through {\em symmetric multiport beam
splitters} (see Ref.~\cite{M.Zukowski:PRA:2564} and references
therein), we have obtained some lower bounds on the maximal violation
of $I_d^+$ with $d\le13$ for the $d$-dimensional maximally entangled
state $\ket{\Psi_d^+}=\frac{1}{\sqrt{d}}\sum_{i=0}^{d-1}
\ket{i_\A}\ket{i_\B}$, where $\ket{i_\A}$ and $\ket{i_\B}$ are,
respectively, the $\idx{i}$ computational basis state of Alice and
Bob's subsystem. Explicit value of these quantum-mechanical violation
for $\ket{\Psi_d^+}$, which we will denote by
$\SqmBIMP{$I_d^+$}(\ket{\Psi^+_d})$, can be found in
Table~\ref{tbl:MaxQuanViolation}. Note that except for $d=5$, these
values also represent the best violation by $\ket{\Psi^+_d}$ that we
were able to find. Moreover, for $d=3$ and 7,
$\SqmBIMP{$I_d^+$}(\ket{\Psi^+_d})$ are in fact the largest quantum
violation of $I_d^+$ that we have found.

\begin{table}[h!]
\caption{\label{tbl:MaxQuanViolation} Bounds on the maximal quantum
violation of $I^+_d$.  The first two columns of the table give the
parameter $d$ and the respective classical upper bound. The next two
columns give the best violation of $I_d^+$ that we have found using
the $d$-dimensional maximally entangled state $\ket{\Psi_d^+}$ in
conjunction with, respectively, the subset of symmetric multiport
measurements~\cite{M.Zukowski:PRA:2564} and arbitrary POVMs. The
fifth and sixth columns of the table give, respectively, the best
lower bound (LB) and the best upper bound (UB) on the maximal quantum
violation of $I^+_d$ that we were able to find. The highest level
semidefinite relaxation~\cite{A.C.Doherty:QMP,
M.Navascues:PRL:010401, M.Navascues:Hierarchy} that we have used to
obtain the UB is listed in the last column of the table.}
\begin{ruledtabular}
\begin{tabular}{r|c|cc|ccc}
$d$ & $\S_{\mbox{\tiny LHV}}$ & $\SqmBIMP{$I_d^+$}(\ket{\Psi^+_d})$
& $\SqmBI{$I_d^+$}(\ket{\Psi^+_d})$ & LB & UB & Level
\\ \hline
  3 &  0.6667  & 0.7124 & 0.7124 & 0.7124 & 0.7124 & 2+ \\
  5 & 0.4800  & 0.5366 & 0.5375 & 0.5376 & 0.5578 & 1 \\
  7 & 0.3878  & 0.4587 & 0.4587 & 0.4587 & 0.4668 & 1 \\
  11 & 0.3058 & 0.3325 & 0.3325 & 0.3328 & - & - \\
  13 & 0.2781 & 0.2987 & 0.2987 & 0.2991  & - & - \\
\end{tabular}
\end{ruledtabular}
\end{table}

For other values of $d$ with $d\le13$, we have nonetheless found larger
quantum violation of $I_d^+$ by combining the iterative method
described in Ref.~\cite{K.F.Pal:PRA:022120} in conjunction with the
lower bound (LB) algorithm introduced in
Ref.~\cite{Y.C.Liang:PRA:2007}. Specifically, the following steps were
repeated a number of times to obtain a non-trivial lower bound on the
maximal quantum violation of $I_d^+$:
\renewcommand{\labelenumi}{(\arabic{enumi})}
\begin{enumerate}
\item Generate alternatively between (i) $\ket{\Psi_d^+}$
    and (ii) a random bipartite pure entangled state in
    $\mathbb{C}^d\ten\mathbb{C}^d$;
\item Find the best violation and hence the optimal
    measurements (for the generated state) using the LB
    algorithm~\cite{Y.C.Liang:PRA:2007, Y.C.Liang:Thesis};
\item Construct the Bell operator $\Bell$ from the
    measurement operators obtained in (2) and determine the
    best violation possible (for these measurements) by
    computing the largest eigenvalue of $\B$;
\item Find the best violation and hence the optimal
    measurements for the eigenstate~\cite{fn:MaxEigState}
    corresponding to the largest eigenvalue obtained in
    (3);
\item Repeat steps (3) and (4) until the best violation found
    converges to the desired numerical precision.
\end{enumerate}
Explicit value of these lower bounds for $d\le13$ can be found in
Table~\ref{tbl:MaxQuanViolation} (see
Appendix~\ref{App:QuantumStategies} for the quantum strategies that
realize some of these violations). From the table, it is clear that
$I^+_d$ and hence $I^+_{c,d}$ [cf. Eq~\eqref{Eq:QuantumBellValues}] for
$d\le 13$ can be violated by quantum mechanics using only projective
measurements and $\ket{\Psi_d^+}$. This is to be contrasted with the
results presented by Ji {\em et~al.}~\cite{Ji:PRA:052103} where they
did not find any legitimate quantum violation of their inequalities for
$d=7$, 11 and 13 using mutually unbiased measurements. For smaller
values of $d$, it is worth noting that our best quantum violation for
$d=3$ agrees with that presented in Ref.~\cite{Ji:PRA:052103}, but for
$d=5$, the best quantum violation that we found is about 2.6\% stronger
than the one presented thereof.

Also included in Table~\ref{tbl:MaxQuanViolation} are upper bounds on
the maximal violation of $I^+_d$ obtained using the semidefinite
relaxation techniques discussed in
Refs.~\cite{M.Navascues:PRL:010401}, \cite{M.Navascues:Hierarchy}
and~\cite{A.C.Doherty:QMP}. Of particular significance is the upper
bound presented for $d=3$, obtained by considering all the second
level operators in the hierarchy introduced in
Refs.~\cite{M.Navascues:PRL:010401,M.Navascues:Hierarchy} plus all
operators of the form $A_1^2B_{\sB}^{\oB}B_{\sB'}^{\oB'}$,
$A_2^1B_{\sB}^{\oB}B_{\sB'}^{\oB'}$,
$A_2^2B_{\sB}^{\oB}B_{\sB'}^{\oB'}$. This upper bound matches exactly
the best lower bound known, thereby proving that the maximal quantum
violation of $I^+_3$ can be obtained using mutually unbiased
measurements~\cite{Ji:PRA:052103}. However, we do not know whether
the upper bound obtained for $d=5$ and $d=7$ can be saturated using
quantum strategies.

\section{Conclusion}
\label{Sec:Conclusion}

In this paper, we have reexamined the class of bipartite, $d$-setting,
$d$-outcome Bell correlation inequalities proposed in
Ref.~\cite{Ji:PRA:052103}. When rewritten in terms of joint
probabilities, we show that the corresponding Bell inequalities for
probabilities naturally generalize the classical winning probability of
the CHSH game introduced in Ref.~\cite{R.Cleve:0404076:2004}. These
Bell inequalities for probabilities, denoted by $\Idp$, are thus also
of interest independent of their correlation counterpart.

In establishing these Bell inequalities explicitly, we have found
that for the more general scenarios of prime value $d>5$, the authors
of Ref.~\cite{Ji:PRA:052103} underestimated the actual classical
upper bounds. Although we could determine the actual classical upper
bound for $d\le 13$ and have provided simple algorithms to estimate
them for larger values of $d$, the general problem is left open in
the present research (the closely related problem for a special class
of two-outcome Bell inequalities, namely, the XOR
games~\cite{R.Cleve:0404076:2004}, is known to be nondeterministic
polynomial-time hard (NP hard)~\cite{J.Hastad:ACM:798}).

Computationally, we have investigated the tightness of the probability
inequality $\Idp$ and its correlation counterpart $I_{c,d}^+$ for
$d\le13$. Our results show that most of these inequalities for prime
value of $d$ are facet-inducing for their respective classical
correlation polytopes. However, none of these inequalities for
non-prime value of $d$ is tight (Appendix~\ref{App:non-prime}). In this
regard, another open problem that follows from our observation is
whether for each $d$, $\Idp$ is facet-inducing if and only if
$I_{c,d}^+$ is facet-inducing.

We have also investigated the quantum violations of $\Idp$ and
compared them against those established in Ref.~\cite{Ji:PRA:052103}.
In particular, we prove that once we are equipped with $\Idp$, the
corresponding correlation analogue proposed by Ji {\em et~al.} is no
longer {\em relevant}~\cite{D.Collins:JPA:2004} (in terms of
detecting an entangled states). On the other hand, we do not know if
$\Idp$ are still relevant once we are equipped with the class of
CGLMP inequalities~\cite{D.Collins:PRL:2002}.

In contrast with most other known tight, multiple-outcome Bell
inequalities~\cite{D.Collins:PRL:2002,D.Kaszlikowski:PRA:2002,
A.Acin:PRA:052325, D.L.Deng:0902.3534}, $I_{c,d}^+$ and $\Idp$ are
apparently not always violated by a non-maximally entangled state. In
particular, among the facet-inducing inequalities investigated, the
best quantum violation that we have found for $d=7$ is actually due to
a maximally entangled state.

\begin{acknowledgments}
YCL acknowledges useful discussions with Stefano Pironio, Llu\'{\i}s
Masanes, Stephanie Wehner, Oded Regev, Se-Wan Ji, and Stephen
Bartlett and financial support from the Australian Research Council.
The authors acknowledge helpful comments from an anonymous referee
and Elena Loubenets on an earlier version of this paper.
\end{acknowledgments}

\appendix

\section{$I_{c,d}^+$ and $I_d^+$ for non-prime $d$}\label{App:non-prime}

For completeness, we will also include our results of computational
investigation in relation to the maximum classical value and the
tightness of $I_{c,d}^+$ and $I_d^+$ for non-prime value of $d$ with
$d\le12$ in the following table.
\begin{table}[h!]
\begin{ruledtabular}
\begin{tabular}{r||ccc||ccc||c}
$d$ & $d_p$ & $r^c_+$ & $I_{c,d}^+$ & $d_p$ & $r^p_+$ & $I_d^+$ &
$\DeltaMax$
\\ \hline
   4 &   48 &   8 & non-tight &   168 &   32 & non-tight & 10\\
   6 &  180 & 146 & non-tight &   960 &  908 & non-tight & 18\\
   8 &  448 &  64 & non-tight &  3248 &  448 & non-tight & 30\\
   9 &  648 &  82 & non-tight &  5328 &  676 & non-tight & 36\\
  10 &  900 & 813 & non-tight &  8280 & 8049 & non-tight & 38\\
  12 & 1584 &  48 & non-tight & 17688 &  576 & non-tight & 60
\end{tabular}
\end{ruledtabular}
\end{table}

\section{Classical Strategies}\label{App:ClassicalStrategies}

Here, we will provide examples of extremal classical strategies that
realize the values of $\Delta_{\mbox{\tiny max}}^{(7)}$,
$\Delta_{\mbox{\tiny max}}^{(11)}$ and $\Delta_{\mbox{\tiny
max}}^{(13)}$ presented in Table~\ref{tbl:DeltaMax}. We will adopt the
notation that the $\idx{k}$ entry of the vector $\oA$ represents
$\oA(k)$, with the exception of $\oA(0)$ which is given as the last
entry of the vector; likewise for $\oB$. For $d=7$, we have
\begin{gather*}
\oA=(0~0~0~1~2~5~0),\\
\oB=(0~5~1~6~0~3~0);
\end{gather*}
for $d=11$, we have
\begin{gather*}
\oA=(0 ~0 ~0 ~1 ~0 ~10 ~8 ~2 ~5 ~7 ~0),\\
\oB=(6 ~0 ~7 ~10 ~1 ~5 ~9 ~0 ~2 ~3 ~0);
\end{gather*}
for $d=13$, we have
\begin{gather*}
\oA=(0~0~1 ~0~6~8~11 ~6 ~4 ~0 ~5 ~9 ~0),\\
\oB=(12~6~7 ~0~6~7~11~11 ~0~10 ~4 ~2 ~0).
\end{gather*}

\section{Quantum Strategies}\label{App:QuantumStategies}

In this Appendix, we will provide the Schmidt coefficients $
c^{(d)}=\left(c_1^{(d)}, c_2^{(d)}, \ldots,
c_{d-1}^{(d)},c_0^{(d)}\right)$ of the quantum state that gives rise to
the best violation that we have found in
Sec.~\ref{Sec:QuantumViolation}. The corresponding quantum states can
then be written explicitly through their Schmidt decomposition, $
\ket{\Psi_d}=\sum_{i=0}^{d-1} c_i^{(d)}\ket{i_\A}\ket{i_\B}$. We will
also provide the phase factors needed to achieve
$\SqmBIMP{$I_d^+$}(\ket{\Psi^+_d})$, the best violation of $I_d^+$ that
we were able to find using the $d$-dimensional maximally entangled
state in conjunction with the measurements facilitated by a symmetric
multiport beam splitter. For this kind of measurements, Alice's and
Bob's POVM element can be written, respectively, as $
\POVMA{\sA}{\oA}=\left(U_{\A}^{\sA}\right)^\dagger\,\Pi_{\oA}\,U_\A^{\sA}$
and $
\POVMB{\sB}{\oB}=\left(U_{\B}^{\sB}\right)^\dagger\,\Pi_{\oB}\,U_\B^{\sB}
$ where $\Pi_{\oA}= \ketbra{\oA}$ and the unitary operators are given
by
\begin{subequations}
\begin{align}
    U_\A^{\sA}=\sum_{k,l=0}^{d-1}\frac{1}{\sqrt{d}}~
    {\rm e}^{\ii 2\pi\left(\frac{kl}{d}+\varphi^{\sA}_l\right)} \ket{k_\A}\bra{l_\A},\\
    U_\B^{\sB}=\sum_{k,l=0}^{d-1}\frac{1}{\sqrt{d}}~
    {\rm e}^{\ii 2\pi\left(\frac{kl}{d}+\phi^{\sB}_l\right)} \ket{k_\B}\bra{l_\B}.
\end{align}
\end{subequations}
Note that for each $\sA$ and $\sB$, we can --- without loss of
generality --- always perform the transformations
$\varphi_l^{\sA}\to\varphi_l^{\sA}-\varphi_0^{\sA}$,
$\phi_l^{\sB}\to\phi_l^{\sB}-\phi_0^{\sB}$ to make
$\varphi_0^{\sA}=\phi_0^{\sB}=0$ for all $\sA$ and $\sB$ while leaving
all the joint probabilities $\ProbTwGJ$ unchanged. This is the
convention that we are going to adopt. In practice, the best multiport
measurements that we have found are those such that the phases for
Alice's and Bob's measurements are equal. In what follows, we will thus
only provide the non-trivial phase factors
\begin{equation}\label{Eq:varphi}
    \varphi^{\sA}=\left(\varphi_1^{\sA},\varphi_2^{\sA},\ldots,
    \varphi_{d-1}^{\sA}\right),
\end{equation}
with the understanding that $\phi^{\sA}=\varphi^{\sA}$ for all
$\sA$.

Explicitly, for $d=5$, the optimal state is
\begin{equation*}
    c^{(5)}=(0.45367, ~0.45049, ~0.44898, ~0.44378, ~0.43899).
\end{equation*}
The corresponding measurements that give rise to the best violation of
$I_5^+$ are non-degenerate, and consist only of rank-1 projectors. On
the other hand, the phase factors needed to achieve $\S_{\mbox{\tiny
QM, mp}}^{I_5^+}\left(\ket{\Psi^+_5}\right)$ are found to be
\begin{equation*}
\begin{array}{ccccccc}
\varphi^{1}=&(0.92207,& 0.65271,& 0.79652,& 0.22729),\\
\varphi^{2}=&(0.34126,& 0.88988,& 0.23114,& 0.36557),\\
\varphi^{3}=&(0.94381,& 0.47445,& 0.19652,& 0.26924),\\
\varphi^{4}=&(0.14166,& 0.23868,& 0.79915,& 0.05082),\\
\varphi^{0}=&(0.96047,& 0.45749,& 0.99915,& 0.84833).\\
\end{array}
\end{equation*}

In the case of $d=7$, the best violation given in
Table~\ref{tbl:MaxQuanViolation} can be achieved using
$\ket{\Psi^+_7}$ together with:
\begin{equation*}
\begin{array}{ccccccccc}
\varphi^{1}=\frac{1}{14}\!\!&( 8,& ~ 7,& ~10,& ~ 2,& ~10,& ~12),\\
\varphi^{2}=\frac{1}{14}\!\!&(12,& ~ 3,& ~ 0,& ~ 2,& ~ 8,& ~10),\\
\varphi^{3}=\frac{1}{14}\!\!&( 0,& ~ 9,& ~12,& ~ 8,& ~10,& ~10),\\
\varphi^{4}=\frac{1}{14}\!\!&( 0,& ~11,& ~ 4,& ~ 6,& ~ 2,& ~12),\\
\varphi^{5}=\frac{1}{14}\!\!&(12,& ~ 9,& ~ 4,& ~10,& ~12,& ~ 2),\\
\varphi^{6}=\frac{1}{14}\!\!&( 8,& ~ 3,& ~12,& ~ 6,& ~12,& ~ 8),\\
\varphi^{0}=\frac{1}{14}\!\!&( 2,& ~ 7,& ~ 0,& ~ 8,& ~ 2,& ~ 2).
\end{array}
\end{equation*}

For $d=11$, and 13, it is expedient to decompose the optimal phase
factors as $\varphi^{\sA}=\varphi^{\sA}_D+\varphi_\epsilon$, where
$\varphi^{\sA}_D$ and $\varphi_\epsilon$ are themselves vectors with
$(d-1)$ entries [cf. Eq.~\eqref{Eq:varphi}]. Explicitly, we have, for
$d=11$,
\begin{equation*}
\begin{array}{ccccccccccc}
\varphi_D^{1}=\frac{1}{22}\!\!\!&( 0,&  0,& 10,& 16,& 16,& 16,&  5,&  0,& 21,&  8),\\
\varphi_D^{2}=\frac{1}{22}\!\!\!&(20,& 18,&  2,&  2,& 20,& 10,&  5,&  8,& 13,&  8),\\
\varphi_D^{3}=\frac{1}{22}\!\!\!&(16,& 10,& 10,&  2,& 14,& 14,& 13,&  0,&  9,& 10),\\
\varphi_D^{4}=\frac{1}{22}\!\!\!&(10,& 20,& 12,& 16,& 20,&  6,&  7,& 20,&  9,& 14),\\
\varphi_D^{5}=\frac{1}{22}\!\!\!&( 2,&  4,&  8,&  0,& 16,&  8,&  9,&  2,& 13,& 20),\\
\varphi_D^{6}=\frac{1}{22}\!\!\!&(14,&  6,& 20,& 20,&  2,& 20,& 19,& 12,& 21,&  6),\\
\varphi_D^{7}=\frac{1}{22}\!\!\!&( 2,&  4,&  4,& 10,&  0,& 20,& 15,&  6,& 11,& 16),\\
\varphi_D^{8}=\frac{1}{22}\!\!\!&(10,& 20,&  4,& 14,& 10,&  8,& 19,&  6,&  5,&  6),\\
\varphi_D^{9}=\frac{1}{22}\!\!\!&(16,& 10,& 20,& 10,& 10,&  6,&  9,& 12,&  3,& 20),\\
\varphi_D^{10}=\frac{1}{22}\!\!\!&(20,& 18,&  8,& 20,&  0,& 14,&  7,&  2,&  5,& 14),\\
\varphi_D^{0}=\frac{1}{22}\!\!\!&( 0,&  0,& 12,&  0,&  2,& 10,& 13,& 20,& 11,& 10),
\end{array}
\end{equation*}
\begin{align*}
    \varphi_\epsilon=\frac{1}{22}(&0.74797, 0.65473, 0.62522, 0.73621, 0.82604,\\
                      &0.02359, 0.36323, 0.92062, 0.82621, 0.36885),
\end{align*}
and for $d=13$,
\begin{equation*}
\begin{array}{ccccccccccccc}
\varphi_D^{1} =\frac{1}{26}\!\!\!&( 1,&\! 21,&\!  1,&\! 25,&\!  3,&\!  5,&\!  2,&\! 18,&\!  7,&\!  3,&\!  9,&\! 10),\\
\varphi_D^{2} =\frac{1}{26}\!\!\!&( 3,&\! 11,&\!  9,&\! 15,&\! 19,&\! 25,&\!  4,&\! 18,&\! 23,&\! 13,&\!  7,&\! 10),\\
\varphi_D^{3} =\frac{1}{26}\!\!\!&( 3,&\! 23,&\! 11,&\! 23,&\! 25,&\!  7,&\! 18,&\!  2,&\! 21,&\!  3,&\!  9,&\! 12),\\
\varphi_D^{4} =\frac{1}{26}\!\!\!&( 1,&\!  5,&\!  7,&\! 23,&\! 21,&\!  3,&\! 18,&\! 22,&\!  1,&\! 25,&\! 15,&\! 16),\\
\varphi_D^{5} =\frac{1}{26}\!\!\!&(23,&\!  9,&\! 23,&\! 15,&\!  7,&\! 13,&\!  4,&\!  0,&\! 15,&\!  1,&\! 25,&\! 22),\\
\varphi_D^{6} =\frac{1}{26}\!\!\!&(17,&\!  9,&\!  7,&\! 25,&\!  9,&\! 11,&\!  2,&\! 14,&\! 11,&\!  9,&\! 13,&\!  4),\\
\varphi_D^{7} =\frac{1}{26}\!\!\!&( 9,&\!  5,&\! 11,&\!  1,&\!  1,&\! 23,&\! 12,&\! 12,&\! 15,&\! 23,&\!  5,&\! 14),\\
\varphi_D^{8} =\frac{1}{26}\!\!\!&(25,&\! 23,&\!  9,&\! 21,&\!  9,&\! 23,&\!  8,&\! 20,&\!  1,&\! 17,&\!  1,&\!  0),\\
\varphi_D^{9} =\frac{1}{26}\!\!\!&(13,&\! 11,&\!  1,&\!  7,&\!  7,&\! 11,&\! 16,&\! 12,&\! 21,&\! 17,&\!  1,&\! 14),\\
\varphi_D^{10}=\frac{1}{26}\!\!\!&(25,&\! 21,&\! 13,&\! 11,&\! 21,&\! 13,&\! 10,&\! 14,&\! 23,&\! 23,&\!  5,&\!  4),\\
\varphi_D^{11}=\frac{1}{26}\!\!\!&( 9,&\!  1,&\! 19,&\!  7,&\! 25,&\!  3,&\! 16,&\!  0,&\!  7,&\!  9,&\! 13,&\! 22),\\
\varphi_D^{12}=\frac{1}{26}\!\!\!&(17,&\!  3,&\! 19,&\! 21,&\! 19,&\!  7,&\!  8,&\! 22,&\! 25,&\!  1,&\! 25,&\! 16),\\
\varphi_D^{0} =\frac{1}{26}\!\!\!&(23,&\!  1,&\! 13,&\!  1,&\!  3,&\! 25,&\! 12,&\!  2,&\! 25,&\! 25,&\! 15,&\! 12),
\end{array}
\end{equation*}
\begin{align*}
    \varphi_\epsilon=\frac{1}{26}(&0.26436, 0.24549, 0.26436, 0.93681, 0.24549,\\
                      &0.24549, 0.84021, 0.84020, 0.26436, 0.93681,\\
                      &0.84020, 0.93681).
\end{align*}

In these two cases, the quantum states that give rise to the
best violation that we have found can be obtained by
determining the eigenvector corresponding to the maximal
eigenvalue of the respective Bell operator $\Bell$. Explicitly,
these quantum states admit the following Schmidt coefficients:
\begin{align*}
    c^{(11)}=(&0.31463,~0.31456,~0.31352,~0.30525,~0.30462,\\
              &0.30432,~0.30116,~0.29086,~0.29048,~0.28915,\\
              &0.28618)\\
    c^{(13)}=(&0.29189,~0.29189,~0.29189,~0.27790,~0.27790,\\
              &0.27790,~0.27502,~0.27502,~0.27502,~0.27329,\\
              &0.27329,~0.27329,~0.24849).
\end{align*}



\begin{thebibliography}{99}

\bibitem{J.S.Bell:1964} J.~S. Bell, Physics (Long Island City,
    N.Y.)  {\bf 1}, 195 (1964).

\bibitem{R.F.Werner:QIC:2001} R.~F.~Werner and M.~M.~Wolf, \qic
    {\bf 1} (3), 1 (2001).

\bibitem{N.D.Mermin:PhysToday:1994} N.~D.~Mermin, Phys. Today
    {\bf 38}, 41 (1985).

\bibitem{A.K.Ekert:PRL:1991} A.~K.~Ekert, \prl {\bf 67}, 661 (1991).

\bibitem{QKD:Security} V.~Scarani and N.~Gisin, \prl {\bf 87},
    117901 (2001); A.~Ac\'in, N.~Gisin, and Ll.~Masanes, \ibid
    {\bf 97}, 120405 (2006).

\bibitem{C.Brukner:PRL:2004} \v{C}.~Brukner, M.~\.Zukowski,
    J.-W.~Pan, and A.~Zeilinger, \prl {\bf 92}, 127901 (2004).

\bibitem{R.Cleve:0404076:2004} R.~Cleve, P.~H\o yer, B.~Toner,
    and J.~Watrous, in {\em Proceedings of the 19th IEEE Annual
    Conference on  Computational Complexity 2004, Amherst, MA},
    (IEEE, New York, 2004), pp. 236-249.

\bibitem{A.C.Doherty:QMP} A.~C.~Doherty, Y.-C.~Liang, B.~Toner,
    and S.~Wehner, in {\em Proceedings of the 23rd IEEE Conference
    on Computational Complexity 2008}, (IEEE Computer Society,
    College Park, MD, 2008), pp. 199-210; see also e-print
    arXiv:0803.4373 (2008).

\bibitem{J.Silman:PLA:3796} J.~Silman, S.~Machnes, and
    N.~Aharon, \pl A {\bf 372}, 3796 (2008).

\bibitem{CHSH}J.~F.~Clauser, M.~A.~Horne, A.~Shimony and
    R.~Holt, \prl {\bf 23}, 880 (1969).

\bibitem{J.S.Bell:Speakable} J.~S.~Bell, {\em Speakable and
    Unspeakable in Quantum Mechanics} (Cambridge University
    Press, Cambridge, England, 2004).

\bibitem{CH:PRD:1974} J.~F.~Clauser and M.~A.~Horne, \prd {\bf 10},
    526 (1974).

\bibitem{fn:two-outcome} Strictly, of course, a correlation
    inequality such as the Bell-CHSH inequality is also
    applicable to an experimental scenario involving an
    arbitrary number of measurement outcomes that are all
    bounded between $-1$ and 1 (see, for example,
    Ref.~\cite{E.R.Loubenets:JPA:455304}).

\bibitem{E.R.Loubenets:JPA:455304} E.~R.~Loubenets, J.~Phys.~A,
    {\bf 41}, 445304 (2008).

\bibitem{D.Collins:JPA:2004} D.~Collins and N.~Gisin,
    J.~Phys.~A {\bf 37}, 1775  (2004).

\bibitem{T.Ito:PRA:2006} T.~Ito, H.~Imai, and D.~Avis, \pra
    {\bf 73}, 042109 (2006).

\bibitem{N.Brunner:PLA:2008} N. Brunner and N. Gisin, Phys.
    Lett. A {\bf 372}, 3162 (2008).

\bibitem{Y.C.Liang:Thesis} Y.-C.~Liang, Ph.D Thesis, University of
    Queensland, 2008; see also e-print arXiv:0810.5400 (2008).

\bibitem{K.F.Pal:PRA:022120} K.~F.~P\'al and T.~V\'ertesi, \pra
    {\bf 79}, 022120 (2009).

\bibitem{Book:MikeIke} M. A. Nielsen and I. L. Chuang, {\it
    Quantum Computation and Quantum Information} (Cambridge
    University Press, Cambridge, England, 2000).

\bibitem{T.C.Ralph:PRA:022313} T.~C.~Ralph, K.~J.~Resch,
    and A.~Gilchrist, \pra {\bf 75}, 022313 (2007).

\bibitem{B.P.Lanyon:NatPhys:134} B.~P.~Lanyon, M.~Barbieri,
    M.~P.~Almeida, T.~Jennewein, T.~C.~Ralph, K.~J.~Resch,
    G.~J.~Pryde, J.~L.~O'Brien, A.~Gilchrist, and A.~G.~White,
    Nat.~Phys. {\bf 5}, 134 (2009).

\bibitem{D.Collins:PRL:2002} D.~Collins, N.~Gisin, N.~Linden,
    S.~Massar, and S.~Popescu, \prl {\bf 88}, 040404 (2002).

\bibitem{D.Kaszlikowski:PRA:2002} D.~Kaszlikowski, L.~C.~Kwek,
    J.-L.~Chen, M.~\.Zukowski, and C.~H.~Oh, \pra {\bf 65},
    032118 (2002).

\bibitem{S.Zohren:PRL:2008} S.~Zohren and R.~D.~Gill, \prl {\bf
    100}, 120406 (2008).

\bibitem{Ll.Masanes:QIC:2003} Ll.~Masanes, \qic {\bf 3}, 345 (2003).

\bibitem{CorPolytope} I.~Pitowsky, {\em Quantum Probability ---
    Quantum Logic} (Springer, Berlin, 1989); Math. Program. {\bf
    50}, 395 (1991); A.~Peres, Found.~Phys. {\bf 29}, 589 (1999).

\bibitem{Polytopes} G.~M.~Zigler, {\em Lectures on Polytopes}
    (Springer, New York, 1995);  B.~Gr\"unbaum, {\em Convex
    Polytopes} (Springer, New York, 2003).

\bibitem{H.Bechmann-Pasquinucci:PRA:2003}
    H.~Bechmann-Pasquinucci and N.~Gisin, \pra {\bf 67}, 062310
    (2003).

\bibitem{A.Acin:PRL:250404} A.~Ac\'in, J.~L.~Chen, N.~Gisin,
    D.~Kaszlikowski, L.~C.~Kwek, C.~H.~Oh, and M.~\.Zukowski,
    \prl {\bf 92}, 250404 (2004).

\bibitem{H.Buhrman:PRA:2005} H.~Buhrman and S.~Massar, \pra
    {\bf 72}, 052103 (2005).

\bibitem{J.L.Chen:PRA78:032107} J.-L.~Chen, C.~Wu, L.~C.~Kwek,
    and C.~H.~Oh, \pra {\bf 78}, 032107 (2008).

\bibitem{S-W.Lee:0803.3097} S.-W.~Lee and D.~Jaksch, \pra {\bf 80},
    010103(R) (2009).

\bibitem{D.L.Deng:0902.3534} D.-L.~Deng, Z.-S. Zhou, and
    J.-L.~Chen, Ann. Phys. {\bf 324}, 1996 (2009).

\bibitem{L.B.Fu:PRL:2004} L.-B.~Fu, \prl {\bf 92}, 130404
    (2004).

\bibitem{W.Son} W.~Son, J.~Lee, and M.~S.~Kim, J.~Phys. A {\bf
    37}, 11897 (2004); \prl {\bf 96}, 060406 (2006).

\bibitem{J.L.Chen:PRA71:032107} J.-L.~Chen, C.~Wu, L.~C.~Kwek,
    D.~Kaszlikowski, M.~\.Zukowski, and C.~H.~Oh, \pra {\bf 71},
    032107 (2005).

\bibitem{Ji:PRA:052103} S.-W.~Ji, J.~Lee, J.~Lim, K.~Nagata,
    and H.-W.~Lee, \pra {\bf 78}, 052103 (2008).

\bibitem{J.L.Chen:PRA:012111} J.-L.~Chen and D.-L.~Deng, \pra
    {\bf 79}, 012111 (2009).

\bibitem{S-W.Lee:PRA:2007} S.-W. Lee, Y.~W.~Cheong, and J.~Lee,
    \pra {\bf 76}, 032108 (2007).

\bibitem{A.Acin:PRA:052325} A.~Ac\'in, T.~Durt, N.~Gisin, and
    J.~I.~Latorre, \pra {\bf 65}, 052325 (2002).

\bibitem{J.L.Chen:PRA:032106} J.-L.~Chen, C.~Wu, L.~C.~Kwek,
    C.~H.~Oh, and M.-L.~Ge, \pra {\bf 74}, 032106 (2006).

\bibitem{N.Gisin:0702021} N. Gisin, in {\em Quantum Reality,
    Relativistic Causality, and Closing the Epistemic Circle: Essays
    in Honour of Abner Shimony}, edited by W.~C.~Myrvold and J.~Christian
    (Springer, The Netherlands, 2009), pp. 125–138; see also e-print
    arXiv:quant-ph/0702021.

\bibitem{S.Popescu:FP:379} S.~Popescu and D.~Rohrlich, \fp {\bf
    24}, 379 (1994).

\bibitem{A.A.Methot:QIC:2007} A.~A.~M\'ethot and V.~Scarani,
    \qic {\bf 7}, 157 (2007).

\bibitem{N.Brunner:NJP:2005} N.~Brunner, N.~Gisin, and
    V.~Scarani, \njp {\bf 7}, 88 (2005).

\bibitem{Ll.Masanes:PRL:2008} Ll.~Masanes, Y.-C.~Liang, and
    A.~C.~Doherty, \prl {\bf 100}, 090403 (2008).

\bibitem{fn:ub} To see this, note that there can be at most $d^2$
    non-zero entries (each upper bounded by 1) in the sum in
    Eq.~\eqref{Eq:BellFn:Prob}.

\bibitem{S.Khot:ACM:2002} S. Khot in {\em Proceedings of the
    34th Annual ACM Symposium on Theory of Computing} (ACM, New
    York, 2002), pp. 767-775.

\bibitem{fn:AllClassicalStrategy} After taking into account of
    the observations that we have made, an exhaustive search
    would still require us to consider $2d^{d-2}$ distinct
    possibilities.

\bibitem{D.Avis:JPA:2005} D.~Avis, H.~Imai, T.~Ito, and
    Y.~Sasaki, J.~Phys.~A {\bf 38}, 10971 (2005).

\bibitem{B.S.Cirelson:LMP:1980} B.~S.~Cirel'son,  Lett. Math.
    Phys. {\bf 4}, 93 (1980).

\bibitem{S.L.Braunstein:PRL:1992} S.~L.~Braunstein, A.~Mann,
    and M.~Revzen, \prl {\bf 68}, 3259 (1992).

\bibitem{fn:POVM} While the non-Hermiticy of the Bell operator can be
    remedied by explicitly including also the complex conjugate of the
    correlation Bell function, such as that done in Eq.~(1) of
    Ref.~\cite{W.Son}, the problem with measuring $(A_{\sA})^n$ and
    $(B_{\sB})^n$ in a way different from $A_{\sA}$ and $B_{\sB}$
    remains, which makes it dubious in calling $I_{c,d}^+$ a
    $d$-setting, instead of a $d(d-1)$-setting Bell correlation
    inequalities.

\bibitem{M.Zukowski:PRA:2564} M.~\.Zukowski, A.~Zeilinger, and
    M.~Horne, \pra {\bf 55}, 2564 (1997).

\bibitem{M.Navascues:PRL:010401} M.~Navascu\'es, S.~Pironio,
    and A.~Ac\'in, \prl {\bf 98}, 010401 (2007).

\bibitem{M.Navascues:Hierarchy} M.~Navascu\'es, S.~Pironio,
    and A.~Ac\'in, New J.~Phys. {\bf 10},
    073013 (2008); S.~Pironio, M.~Navascu\'es,
    and A.~Ac\'in, e-print arXiv:0903.4368 (2009).

\bibitem{Y.C.Liang:PRA:2007} Y.-C.~Liang and A.~C.~Doherty,
    \pra {\bf 75}, 042103 (2007).

\bibitem{fn:MaxEigState} The eigenstate of $\Bell$ giving the
    largest eigenvalue is the quantum state that gives the
    highest violation of $I_d^+$ for the measurements defining
    $\Bell$.

\bibitem{J.Hastad:ACM:798} J.~H{\aa}stad, J. ACM, {\bf 48},
    798 (2001).

\end{thebibliography}
\end{document}